%% file: Final_Report.tex
\documentclass{SPhdThesis}

\usepackage{amsthm,amsmath,amssymb}
\usepackage{mathrsfs}
\usepackage{graphics}
\usepackage{amsfonts}
\usepackage{amssymb}
\usepackage{amsbsy}
\usepackage{latexsym}
\usepackage{epsfig}
\usepackage{fullpage}
\RequirePackage[numbers]{natbib}
\usepackage[latin1]{inputenc}
\usepackage{algorithm}
\usepackage{algorithmic}
\usepackage[T1]{fontenc}
\usepackage{lmodern}
\usepackage[export]{adjustbox}
\usepackage[toc,page]{appendix}
\usepackage{graphicx}

\renewcommand{\phi}{\varphi}
\renewcommand{\epsilon}{\varepsilon}

\SgSetTitle{Dynamic Mean-Variance Portfolio Optimisation}
\SgSetAuthor{Meng Xiang}
\SgSetSupervisor{Assistant Prof. Zhou Chao}
\SgSetYear{2017/2018}
\SgSetDepartment{Department of Mathematics}
\SgSetUniversity{National University of Singapore}
\SgSetDeclarationDate{April 2018}

\begin{document}
	\begin{frontmatter}
		\SgAddTitle%
		\input{acknowledgments}%
		\input{abstract}%
		\SgAddToc
	\end{frontmatter}

\chapter{Introduction}

In modern mathematical finance, portfolio optimisation has been a fundamental and central topic to understanding the stock market and making decisions. It was Harry Markowitz that made the major breakthough of the area in 1952 \cite{markowitz}, whose theory is commonly known as Modern Portfolio Theory (MPT). The theory provides an answer to the fundamental question of how should an investor allocate funds among all possible combinations of assets. All theories before Markowitz had obstacled by the trade-off between the return and risk of investments. Markowitz quantified the return and risk using the statistical measure of sample mean and variance, and further showed that both return and risk should be considered together in order to make an optimal selection.\\\\
There are a few reasons why the theory made a revolutionary change in investment decisions. Firstly, the theory shows that the risk of a portfolio does not only depend on the risk of individual constituents, but also relies on their co-movements. A qualitative principle corresponding to this is the diversification of unsystematic risks. In the classical point of view, investors aim to select securities that can generate the highest future cash flow given the current price. However, Markowitz's portfolio theory suggests that the correlation among the constituents in a portfolio makes an important role in deciding the total riskiness, which is another factor that should be considered in decisions. Secondly, the theory formulated the decision process as an optimisation problem. It suggests that an optimal, or efficient portfolio should be the one that has the minimum variance given a certain level of expected return. Any other combinations of securities are considered as inefficient portfolios because they have larger variance. In other words, we are looking for the portfolio that possesses least risk while retaining the same return.\\\\
Since Markowitz's initial publication, there has been plenty of works contributing to development of the mean-variance optimisation (MVO) so far. In this paper, one kind of them, namely, dynamic mean-variance optimisation (DMVO) is mainly discussed. The traditional MVO is considered as a static strategy, in the sense that it optimise the portfolio only at each time point. In comparison, DMVO seeks for a strategy that can optimise the wealth over both each instant interval and the whole investment horizon. One major obstacle for DMVO is that the Bellman Optimality Principle does not hold in a multiperiod setting of MVO. That is, the problem is time inconsistent under DMVO objective. Hence, the word "optimal" is not clear on the object that it refers to, and the dynamic programming cannot be directly applied.\\\\
Among research that focuses on DMVO, people normally handle the time-inconsistency with two approaches. The first method is to study the precommitment problem. In this case, the term "optimal" is interpreted as the optimum of the terminal wealth viewed from the starting time. Richardson (1989) \cite{richardson} turns out to be the pioneer that studies the precommitment problem in a countinuous-time setting. For the discrete-time case, Li and Ng (2000) \cite{ling} came out with an embedding technique. They converted the time-inconsistent problem into a stochastic linear quadratic (LQ) control problem.\\\\
The second resolution is to adapt a game-theoritical view, which is the main topic of the thesis. Notice that the essence of time-inconsistency is that the investor can deviate from the future optimal policy to obtain a optimum in next instant interval. One can take incentives to deviate the current policy into acocunt, and perceive the future preferences as players. The objective is then redirected to finding the Nash equilibrium for the game. This game-theoritical approach was firstly studied by Strotz (1955 \cite{strotz} where a deterministic Ramsey problem was studied. In terms of DMVO, Basak and Chabakauri (2010) \cite{basak} are first authors to tackle the problem using game theoritical approach. The authors cleverly adjust the time-inconsistent DMVO to a time-consistent objective using the total variance formula. After that, they manage to apply Hamilton-Jacobi-Bellman (HJB) equation to solve the explicit solution, under a series of Brownian Motion (BM) driven dynamics. Bjork and Murgoci (2009) \cite{bjork} extend the idea in more general terms. The approach, which studies a general form of dynamic objective function, is able to handle different market settings and a certain group of stochastic control problems. Bjork, Murgoci and Zhou \cite{bjorkzhou} further study the case of general risk aversion in 2014. In their research, risk aversion is treated as the function of wealth instead of a constant, and a linear system solution is derived using dynamic programming. They also derived the explicit solution to a specific case, studied its practical behaviour, and proved it is economically reasonable.\\\\
The paper is organised as follows: In section 2, we review the classical MVO problem. In section 3, we study a time-consistent strategy of DMVO. In section 4, we incorporate the theoritical result with market data. Two set of market data, one from the real world and the other from simulation, are chosen as samples to illustrate effectiveness of our strategies. In section 5, we compare results under our "theoritical" strategies to those from the Long Short-Term Memory (LSTM), an representative apporach where the deep learning technique is applied.

\newpage
\chapter{Mean-Variance Optimisation}
Assume the stock market consists $n$ risky assets $S_1, S_2, ... , S_n$. Denote their means returns as the vector $ \pmb{\mu}= (\mu_1, \mu_2, . . . , \mu_n)$ and the covariance matrix as $\Sigma = (\sigma_{ij})$. An investment portfolio is represented by the vector $\pmb{\omega} = (\omega_1, \omega_2, . . .,\omega_n)$, where each $\omega_i$ represents the proportion of the money invested in the stock $i$. The portfolio mean and variance, which are proxies for portfolio return and risk, is given by $\mu = \pmb{\omega^T} \pmb{\mu}$ and $\sigma^2 = \pmb{\omega^T}\Sigma \pmb{\omega}$ respectively. In the paper, we assume that the covariance matrix is positive-definite, i.e. $\pmb{\omega^T}\Sigma \pmb{\omega}>0$ for every $\pmb{\omega} \neq \pmb{0}$. This constraint ensures that each strategy made up by $S_1, S_2, ... , S_n$ cannot be replicated by the others. \\\\
Given a certain level of expected portfolio return, one seeks a portfolio $\pmb{\omega}$ that minimises the risk. The MVO problem takes the form:
\begin{align}
min \: \frac{1}{2} \pmb{\omega^T}\Sigma \pmb{\omega}&\\
s.t. \quad \pmb{1^T \omega} = 1&\\
\pmb{\omega^T} \pmb{\mu} \geq \mu&
\end{align}
This formulation satisfies the normal investment case. For a certain market, a portfolio manager could have lowest expected return in mind, based on the market feature and global trend. However, the risk does not seem to be obvious. It is then natural to define the objective of the strategy as minimising the portfolio risk. Other formulations are readily available on literatures, for example, to maximize the expected return subject to an upper limit on the portfolio risk.\\\\
We observe for the problem above, if we set the last constraint to equality, the MVO will then becom a quadratic problem which can be solved analytically using method of Lagrange multiplier. In this case, the problem is formulated as
\begin{align}
min \: \frac{1}{2} \pmb{\omega^T}\Sigma \pmb{\omega} \label{mvo}&\\
s.t. \: \pmb{1^T \omega} = 1 \label{constraint1}&\\
\pmb{\omega^T} \pmb{\mu} = \mu \label{constraint2}&
\end{align}
The small adjustment changes a soft constraint to a binding constraint, which makes the analytical solution possible. However, it deviates from the realilty because no portfolio manager requires an exact level of the return (in fact, more return is always better). This is also the reason why the strategy is rarely utilized by investment managers directly. Nevertheless, we will see in Chapter 4 that the strategy still provides some enlightenment to us.\\\\
To solve to problem, we firstly obtain the Lagrangian:
\begin{equation} \label{lagrangian}
L(\pmb{\omega}, \lambda_1, \lambda_2) = \frac{1}{2} \pmb{\omega^T}\Sigma \pmb{\omega}- \lambda_1 (\pmb{1^T \omega}-1) - \lambda_2 ( \pmb{\omega^T} \pmb{\mu}-\mu)
\end{equation}
Then set the first-order derivative of \ref{lagrangian} to zero:
\begin{equation}
\Sigma \pmb{\omega}- \lambda_1 - \lambda_2 \pmb{\mu}=0
\end{equation} 
Taking \ref{constraint1} and \ref{constraint2} into account, our focus is now to solve the linear system
\begin{eqnarray}
\pmb{\omega}- \lambda_1 - \lambda_2 \pmb{\mu}=0\nonumber\\
\pmb{1^T \omega} = 1 \nonumber \\
\pmb{\omega^T} \pmb{\mu} = \mu 
\end{eqnarray}
Define the following constants
\begin{align*}
a = \pmb{1^T} \Sigma \pmb{1} \quad b = \pmb{1^T} \Sigma \pmb{\mu} \quad
c = \pmb{\mu^T} \Sigma \pmb{\mu} 
\end{align*}
Applying Cauchy-Schwartz inequality, and noting that the covariance is positive-definite, we can easily see that $ac > b^2$. Hence, the solution of the system, or the optimal portfolio is given by
\begin{equation}
\pmb{\omega^*} = (\frac{c-b\mu}{ac-b^2}) \Sigma^{-1} \pmb{1} + (\frac{a\mu - b}{ac-b^2})\Sigma^{-1} \pmb{\mu}
\end{equation}
From the solution above, we observe that the strategy is highly dependent on mean return, covariance matrix and the expected return. In general, stocks which has a better risk-return ratio (Sharpe ratio) will be allocated with more weight. In addition, we will see in Chapter 4 that the expected return influences the volatility of the investment.

\newpage
\chapter{Dynamic Mean-Variance Optimisation}
\section{Market Assumptions}
We firstly consider a market satisfying Markov property and consisting two assets: a riskless bond and a risky stock. We further assume a finite investment horizon of $[0,T]$. Uncertainty is represented by a filtered probability space $(\Omega, \mathscr{F}, \{\mathscr{F}_t\}, P)$. Two Brownian motions with correlation $\rho$ are defined on the space, namely $w$ and $w_x$. Then, $\{\mathscr{F}_t\,t\in [0,T]\}$ is in fact the augmented filtration generated by $w$ and $w_x$, and we assume all stochastic processes are adapted by it. In addition, all processes are assumed to be well-defined without explicitly stated regularity conditions. Let r be the constant interest rate of the riskless bond, and let the stock price follows the dynamic:
\begin{equation} \label{stock}
dS_t =  \mu S_tdt + \sigma S_tdw_t
\end{equation}
where $\mu$ is the mean return of the stock and $\sigma$ is the volatility. Introduce the state variable X that satisfies
\begin{align} \label{state}
dX_t = m(X_t, t) dt + \nu (X_t, t)dw_{Xt} 
\end{align}
In the following part of the paper, we use $\mu_t, \sigma_t, m_t, \nu_t$ to denote four coefficients defined in \ref{stock} and \ref{state}. It is clear that the market is incomplete when $-1<\rho<1$ since the uncertainties in the state variable cannot be perfectly hedged by any portfolio consisting only the bond and stock. On the other hand, for the special case when $\rho = \pm 1$, the market attains dynamic completeness.\\\\
Assume an investor has initial wealth $W_0$ at time $0$. Define the policy she chooses as $\theta_t$, the amount of the money she invests at time $t$. Under the market setup, the wealth follows the process below:
\begin{align} \label{wealthconstraint}
dW_t = [rW_t + \theta_t(\mu_t - r)]dt + \theta_t \sigma_t dw_t
\end{align}
Assume the investor seek for a dynamic mean-variance strategy with objective to maximize the terminal wealth $W_T$. Instead of the static MVO formulation given by \ref{mvo}, \ref{constraint1} and \ref{constraint2}, we rewrite the Lagrangian \ref{lagrangian} into the following form:
\begin{align}
&\max_{\theta} E[W_T]-\frac{\gamma}{2}var[W_T] \\
s.t. dW_t = &[rW_t + \theta_t(\mu_t - r)]dt + \theta_t \sigma_t dw_t \nonumber
\end{align}
\section{Determination of the DMVO Policy}
The solution given by Basak and Chabakauri (2010) \cite{basak} is based on dynamic programming approach. It is similar to the pioneer approach given by Strotz (1056) \cite{strotz}, which attempts to handle the time-inconsistency by recursive relations. Specifically, the main reason that we cannot apply dynamic programming directly is due to the lack of tower property for the variance term. Here, by tower property, we mean the iterative expectation formula:
\begin{align}
E_t[E_{t+\tau}[W_T]] = E_t[W_T]
\end{align}
This is not a problem for the expectation, but it makes trouble for the variance, especially $E[W_T^2]$ where the tower property cannot be applied. To tackle the issue, we need to change the original DMVO objective slightly. The rescue is provided by the decomposition formula of variance, which is of the following form:
\begin{align}
Var_t[W_T] = E_t[Var_{t+\tau}(W_T)]+Var_t[E_{t+\tau}(W_T)] \label{totalvariance}
\end{align}
This formula also provides information on how optimal investment policy is chosen: Fix a time $t$, the investment policy $\theta_{\tau}$ not only aims to optimise the expected risk of time $t+\tau$, but also accounts for the variance of expected terminal wealth at time $t+\tau$. As a result, at time $t+\tau$, the investor could have incentive to deviate from the optimal policy obtained at time $t$. We recall the DMVO objective function at time t:
\begin{align}
U_t= E_t[W_T]-\frac{\gamma}{2}Var_t[W_T] \label{dmvo1}
\end{align}
We now incorporate the incentive to deviate to policy into the DMVO objective, i.e. substitute \ref{totalvariance} into \ref{dmvo1}. Further applying the tower property to $E[W_T]$, we obtain the new DMVO objective:
\begin{align}
U_t= E_t[U_{t+\tau}]-\frac{\gamma}{2}Var_t[E_{t+\tau}(W_T)] \label{dmvo2}
\end{align}
This adjustment makes it possible for us to determine the optimal policy recursively, and converts the problem to a time-consistent problem. In other words, the investor chooses her optimal policy by considering that the policy will be always optimal in the future and she is able to revise the policy through later in the investment horizon.\\\\
Next, we seek for a recusive relation with respect to the value function, which will be then used to derive the HJB equation. Given the optimal policy $\theta_s^*$, $s\in [t,T]$ derived by backward induction, we define the value function $J_t$ as:
\begin{align}
J_t = J(W_t, S_t, X_t, t) := E_t[W_T^*] - \frac{\gamma}{2}Var_t[W_T^*] 
\end{align}
Note that the optimal terminal wealth $W_T^*$ is deduced by the optimal policy $\theta_s^*$, $s\in [t,T]$. Denote $[t,t+\tau], \tau>0$ as the decision interval, i.e., the investor can opt to adjust her time $t$ investment policy after the interval at time $t+\tau$. We further assume that the investor follows the optimal policy $\theta_s^*$ from $t+\tau$ to $T$. Then, regarding the recursive step, the problem is reduced to seek an investment policy $\theta_s$ from $t$ to $t+\tau$ such that the following objective is maximized:
\begin{align}\label{dmvo3}
E_t[J_{t+\tau}] - \frac{\gamma}{2}Var_t[E_{t+\tau}(W_T)] 
\end{align}
Note again due to existance of the second term (time-consistency adjustment term), the optimal policy $\theta_s^*$ over $[t+\tau,T]$ is not necessarily optimal over $[t,t+\tau]$. \\\\
Before getting to the HJB equation, we would like to rewrite the objective over $[t,t+\tau]$. Applying Ito's Lemma to the wealth constraint \ref{wealthconstraint}, we have:
\begin{align} \label{ito1}
d(W_t e^{r(T-t)}) = \theta_t(\mu_t - r) e^{r(T-t)} dt + \theta_t\sigma_t e^{r(T-t)}dw_t
\end{align}
Integrating \ref{ito1} from $t+\tau$ to $T$, and take the expectation at time $t+\tau$, we obtain:
\begin{align} \label{ito2}
E_{t+\tau}[W_T^*] = W_{t+\tau}e^{r(T-t-\tau)} + f_{t+\tau}
\end{align}
where $W_T^*$ is the optimal wealth followed by $\theta_s^*$ over $[t+\tau,T]$, $f_t = E_t[W_T^*] - W_te^{r(T-t)} = E_t[\int_{t}^{T} \theta_s^*(\mu_s - r)e^{r(T-s)}ds]$ represents anticipated gain over $[t, T]$. Substituting \ref{ito2} to \ref{dmvo3}, and further noticing that $f_t$ and  $W_te^{r(T-t)}$ are adapted to the filtration $\mathscr{F}_t$, the objective over $[t, t+\tau]$ is rewritten as:
\begin{align} 
J_t =\max_{\theta_s, s\in [t, t+\tau]} \{E_t[J_{t+\tau}] - \frac{\gamma}{2} Var_t [f_{t+\tau} - f_t + W_{t+\tau}e^{r(T-t-\tau)} - W_te^{r(T-t)}] \} \label{dmvo4} 
\end{align}
subject to the wealth constraint \ref{wealthconstraint} and the terminal condition $J_T = W_T$. The terminal condition is derived as a result of $Var_T[W_T] = 0$ and $E_T[W_T] = W_T$. It is obvious from \ref{dmvo4} that $f_{t+\tau}$ is also determined by the optimal policy $\theta_s^*$ over $t+\tau$ to $T$. Hence, we establish connection among $\theta^*_t$, $f_t$, $J_t$ by the HJB equation given in the following Lemma:\\
\newtheorem{lemma}{Lemma}
\begin{lemma}
	The value function $J$ under DMVO objective followes the recursive relation
	\begin{align}
	0 = \max_{\theta_t} E_t[dJ_t] - \frac{\gamma}{2} Var_t[df_t+d(W_t e^{r(T-t)})] \label{hjb1}\\
	s.t. \quad d(W_t e^{r(T-t)}) = \theta_t(\mu_t - r) e^{r(T-t)} dt + \theta_t\sigma_t e^{r(T-t)}dw_t \nonumber\\
	J_T = W_T \nonumber
	\end{align}
\end{lemma}
Interestingly, we found from the definition of $f_t$ in \ref{ito2} that $f_t$ is computed via the optimal strategy $\theta^*_t$ and does not depend on $W_t$. Hence, for the HJB equation \ref{hjb1}, $\theta_t$ should have no effect on $df_t$. \\\\
In addition, we notice that $W_t$ does not affect the optimal policy $\theta_t^*$. This is shown by the integrating \ref{ito1} from $t$ to $T$ and take the expection at time $t$:
\begin{align}
J_t = E_t[W_T] -\frac{\gamma}{2} Var_t[W_T] &= W_t e^{r(T-t)} + E_t[\int_{t}^{T} \theta_s(\mu_s - r)e^{r(T-s)}ds]\label{ito3} \\
&- \frac{\gamma}{2} Var_t[\int_{t}^{T} \theta_s(\mu_s - r)e^{r(T-s)}ds + \int_{t}^{T} \theta_s \sigma_s e^{r(T-s)} dw_s] \nonumber
\end{align}
From \ref{ito3}, we notice that the value function $J_t$ is separable in $ W_t e^{r(T-t)} $ and a function that does not depend on $W_t$. As a result, at any time $t\in[0,T]$, the optimal investment policy $\theta_s^*$ does not depend on the current wealth $W_s$ for $s\geq t$, since $\theta_s^*$ is worked out by backward induction. Specifically, we separate $J_t$ into two parts:
\begin{align}
J(W_t, S_t, X_t, t) = W_t e^{r(T-t)}+\tilde{J} (S_t, X_t,t)
\end{align}
Noticing the analysis above, we then apply Ito's Lemma on \ref{hjb1} and obtain the following equation:
\begin{align}
0 = &\max_{\theta_t}\{D\tilde{J_t}dt + \theta_t(\mu_t-r)e^{r(T-t)}dt - \nonumber\\
&\frac{\gamma}{2}Var_t[\sigma_t S_t\frac{\partial f_t}{\partial S_t}dt + \nu_t \frac{\partial f_t}{\partial X_t}dw_Xt+\theta_t \sigma_t e^{r(T-t)}dw_t]\} \nonumber\\
=&\max_{\theta_t}\{D\tilde{J_t}dt + \theta_t(\mu_t-r)e^{r(T-t)}dt - \nonumber\\
&\frac{\gamma}{2}[\sigma_t^2 S_t^2(\frac{\partial f_t}{\partial S_t})^2 + \nu_t^2 (\frac{\partial f_t}{\partial X_t})^2+\theta_t^2 \sigma_t^2 e^{2r(T-t)}+\nonumber\\
&2\rho \nu_t \sigma_t S_t \frac{\partial f_t}{\partial S_t} \frac{\partial f_t}{\partial X_t}+ 2\theta_t \sigma_t (\sigma_t S_t \frac{\partial f_t}{\partial S_t}+\rho \nu_t \frac{\partial f_t}{\partial S_t})e^{r(T-t)}]\} \label{ito4} \\
s.t. &\quad \tilde{J_T} = 0 \nonumber
\end{align}
where $D$ is the operator that denotes Ito's Lemma for functions involved in two processes $S_t$ and $X_t$:
\begin{align}
DF(S_t, X_t, t) = \frac{\partial F_t}{\partial t} + \mu_t S_t \frac{\partial F_t}{\partial S_t} + m_t \frac{\partial F_t}{\partial X_t} +\frac{1}{2} (\sigma_t^2 S_t^2 \frac{\partial^2 F_t}{\partial S_t^2} + \nu_t^2 \frac{\partial^2 F_t}{\partial X_t^2} + 2\rho \nu_t \sigma_t S_t \frac{\partial^2 F_t}{\partial X_t \partial S_t})
\end{align}
It is not surprising from \ref{ito4} that besides the normal term $D\tilde{J_t}dt + \theta_t(\mu_t-r)e^{r(T-t)}$, there are other terms adjusting the investment policy $\theta_t$ and the anticipated gain $f_t$. Furthermore, \ref{ito4} is in fact a maximazation problem on quadratic function of $\theta_t$. Noticing this property, we have a the following proposition:\\
\newtheorem{prop}{Proposition}
\begin{prop}
	The optimal stock investment policy of a dynamic mean-variance optimizer is given by: 
	\begin{align}
	\theta_t^* = \frac{\mu_t-r}{\gamma \sigma^2}e^{-r(T-t)} - (S_t \frac{\partial f_t}{\partial S_t}+\frac{\rho \nu_t}{\sigma_t} \frac{\partial f_t}{\partial S_t})e^{-r(T-t)}  \label{optimalpolicy1}
	\end{align}
	where $f_t = E_t[W_T^*] - W_te^{r(T-t)} = E_t[\int_{t}^{T} \theta_s^*(\mu_s - r)e^{r(T-s)}ds]$ represents the expected total gains or losses from the stock investment\\
\end{prop}
From the expression above, we observe that the optimal policy is made up by two parts, namely myopic demand and hedging demand. Firstly, the myopic demand $\dfrac{\mu_t-r}{\gamma \sigma^2}$ is popularly known as the Sharpe Ratio \cite{sharpe} incorporated with the risk aversion $\gamma$. Typically, it tells the investor how well the return of an asset compensates for the risk taken. Making investment based on the Sharpe Ratio is a strategy that optimises return only for the next small time interval, without considering fluctuations from future investments. That is why we refer it to the name "myopic demand". Secondly, the hedging demand arises as a result of compensating future fluctuations under a dynamic optimisation objective. Moreover, the term $(S_t \dfrac{\partial f_t}{\partial S_t}+\dfrac{\rho \nu_t}{\sigma_t} \dfrac{\partial f_t}{\partial S_t})$ implies it is sensitivities of anticipated portfolio gains to stock prices ($\dfrac{\partial f_t}{\partial S_t}$) and condition of state variables ($\dfrac{\rho \nu_t}{\sigma_t}$) that drive the hedging demand, which gives an intuition about factors that affect the strategy. We can actually show this relationship quantitatively. Recall dynamics of stock and state variables are given by $dS_t =  \mu S_tdt + \sigma S_tdw_t$, $dX_t = m_t dt + \nu_t dw_{Xt} $, and the correlation between $w_t$ and $w_{Xt}$ is $\rho$. We rewrite the hedging term as follows:
\begin{align}
-S_t \frac{\partial f_t}{\partial S_t}+\frac{\rho \nu_t}{\sigma_t} \frac{\partial f_t}{\partial S_t}
&= -\frac{1}{\sigma_t^2 dt} (S_t \sigma_t^2 \frac{\partial f_t}{\partial S_t} dt + \rho \nu_t\sigma_t \frac{\partial f_t}{\partial S_t}dt)\nonumber \\
&= -\frac{1}{\sigma_t^2 dt} (cov(\sigma_t dw_t, S_t \sigma_t \frac{\partial f_t}{\partial S_t} dw_t+\frac{\partial f_t}{\partial X_t}\nu_t dw_{Xt}))\nonumber \\
&= -\frac{1}{\sigma_t^2 dt} cov(\frac{dS_t}{S_t}, df_t)
\end{align}
From the expression above, the hedging demand is positive (negative) when the short-term stock return is negatively (positively) correlated to the short-term anticipated portfolio gain. Assume in a small interval, there is a negative correlation between stock and portfolio gain, which means the stock will move in a opposite direction as the portfolio gain. In this case, investing in the stock will lead to a smaller variance for the portfolio, which is an optimising direction for the mean-variance objective. Hence, the negative correlation leads to a positive hedging demand.\\\\
In proposition 1, we have deduced the expression of the optimal policy. However, \ref{optimalpolicy1}
only gives an implicit formula. Specifically, $f_t = E_t[W_T^*] - W_te^{r(T-t)} = E_t[\int_{t}^{T} \theta_s^*(\mu_s - r)e^{r(T-s)}ds]$ is not a computable formula. Our aim now is to obtain a expression of optimal policy with explicitly characterized market parameters $\mu_t, \sigma_t, m_t, \nu_t$. To solve the problem, we substitute the optimal policy \ref{optimalpolicy1} into the definition of $f_t$:
\begin{align}
f_t =& E_t[\int_{t}^{T} \theta_s^*(\mu_s - r)e^{r(T-s)}ds] \nonumber \\
=& E_t[\int_{t}^{T} \frac{1}{\gamma}(\frac{\mu_s - r}{\sigma_s})^2 ds] - E_t[\int_{t}^{T} (S_s \frac{\partial f_s}{\partial S_s}+\frac{\rho \nu_s}{\sigma_s} \frac{\partial f_s}{\partial S_s})(\mu_s - r)ds]\label{anticipatedgain1}
\end{align}
It is not difficult to observe that the first term comes from the myopic demand and the second term arises from the hedging demand. In order to avoid the second term, which prevents us to do the explicit computation, we now seek for a probabiliy measure that can potentially eliminate the hedging effect. Noticing that $f_t + W_t e^{r(T-t)} = E_t[W_T^*]$ is a martingale, we then apply Ito's Lemma to it and obtain:
\begin{align}
d (f_t + W_t e^{r(T-t)} )=&\frac{\partial f}{\partial t} dt + \frac{\partial f}{\partial S} dS + \frac{\partial f}{\partial X} dX + \frac{1}{2} \frac{\partial^2 f}{\partial S^2} dS^2 + \frac{1}{2} \frac{\partial^2 f}{\partial X^2} dX^2 +  \frac{\partial^2 f}{\partial S \partial X} dS dX \nonumber\\
&+[ \frac{(\mu_t - r)^2}{r \sigma_t^2} - (S_t \frac{\partial f}{\partial S} + \frac{\rho \nu_t }{\sigma_t} \frac{f_t}{X-t}) ] dt \nonumber\\
&+  [ \frac{(\mu_t - r)^2}{r \sigma_t} - (S_t \sigma_t \frac{\partial f}{\partial S} + \rho \nu_t \frac{f_t}{X-t}) ]dw_t
\end{align}
Since $f_t + W_t e^{r(T-t)}$ is a martingale, the $dt$ part of above must be zero, that is:
\begin{align}
\frac{\partial f}{\partial t} &+ rS_t \frac{\partial f}{\partial S}+ (m_t - \rho \nu_t \frac{\mu_t - r}{\sigma_t})\frac{\partial f}{\partial X}\nonumber \\
&+ \frac{1}{2} (\sigma_t^2 S_t^2 \frac{\partial^2 f}{\partial S^2} + \nu_t^2 \frac{\partial^2 f}{\partial X^2} + 2\rho \nu_t \sigma_t S_t \frac{\partial^2 f}{\partial X \partial S})+ \frac{1}{\gamma} (\frac{\mu_t -r}{\sigma_t})^2= 0 \label{feynmankac1}
\end{align}
By Feynman-Kac theorem, \ref{feynmankac1} gives unique solution to the following equation:
\begin{align}
f_t = E_t^*[\int_{t}^{T}\frac{1}{\gamma} (\frac{\mu_s - r}{\sigma_s})^2 ds] \label{anticipatedgain2}
\end{align}
where $E^*_t[\cdot]$ denotes the expectation under the probability measure $P^*$. Under $P^*$, the stock and the state variable adapt the following processes:
\begin{align}
\frac{dS_t}{S_t} &= rdt + \sigma_t dw_t^* \label{stockprocess2}\\
dX_t &= (m_t - \rho \nu_t \frac{\mu_t - r}{\sigma_t})dt + \nu_t dw_{Xt}^* \label{stateprocess2}\\
with \quad dw_t^* &= dw_t + \frac{\mu_t - r}{\sigma_t}dt, \quad dw_{Xt}^* = dw_{Xt} + \rho \frac{\mu_t - r}{\sigma_t}dt \label{newbrownian}
\end{align}
It is exciting that \ref{anticipatedgain2} fully gets rid of the "troublesome" hedging term in \ref{anticipatedgain1}, and completely keeps the myopic term. Therefore, we can refer the new measure $P^*$ as a hedge-neutral measure. Furthermore, if the market is complete ($\rho=\pm 1$), the hedge-neutral measure coincides with the risk-neutral measure. We then summarize finding above as Proposition 2:\\
\begin{prop}
	The optimal investment policy is denoted by:
	\begin{align}
	\theta_t^* = &\frac{\mu_t-r}{\gamma_t \sigma^2}e^{-r(T-t)} - \nonumber\\
	&(S_t \frac{\partial E_t^*[\int_{t}^{T} \frac{1}{\gamma}(\frac{\mu_s - r}{\sigma_s})^2 ds]}{\partial S_t}+\frac{\rho \nu_t}{\sigma_t} \frac{\partial E_t^*[\int_{t}^{T} \frac{1}{\gamma}(\frac{\mu_s - r}{\sigma_s})^2 ds]}{\partial S_t})e^{-r(T-t)} \label{explicitpolicy}
	\end{align}
	where the anticipated portfolio gain is be represented as:
	\begin{align}
	f_t= f(S_t,X_t,t) = E_t^*[\int_{t}^{T}\frac{1}{\gamma} (\frac{\mu_s - r}{\sigma_s})^2 ds]
	\end{align}
	under a hedge-neutral mesure $P^*$, where two standard Brownian motions with correlation $\rho$ are given by:
	\begin{align}
	dw_t^* &= dw_t + \frac{\mu_t - r}{\sigma_t}dt, \quad dw_{Xt}^* = dw_{Xt} + \rho \frac{\mu_t - r}{\sigma_t}dt
	\end{align}
	The Radon-Nikodym derivative from $P$ to $P^*$ is given by:
	\begin{align}
	\frac{dP^*}{dP} = e^{-\frac{1}{2}\int_{0}^{T}(\frac{\mu_s-r}{\sigma_s})^2 ds - \int_{0}^{T}\frac{\mu_s-r}{\sigma_s} dw_s}
	\end{align}
\end{prop}
Equation \ref{explicitpolicy} presents an explicit expression of investment policy, from which all parameters are characterized by the market itself. For situations where analytical solution cannot be obtained, we can go through Monte-Carlo methods to obtain numerical results. It is necessarily to notice that simulations have to be conducted under the hedge-neutral measure $P^*$, and partial derivatives can rewritten as Malliavin derivatives in order to be adapted by Monte-Carlo.

\section{Game-theoretic Interpretation}
We recalled that for the case of DMVO, time-inconsistency is an obstacle that prevents dynamic programming. We have handled the issue using the "magic" formula of total variance. However, this is not a pure coincidence. While deriving the HJB equation, we assume we already have the future optimal strategy and react optimally in the next instant time interval. In this case, choosing optimal strategy is to reach a pure-strategy Nash equilibrium in the game. \\\\
To be more precise, we assume there are infinitely many players standing at each time point on the investment horizon $[0,T]$. Each player takes an investment strategy $\theta_t$ such that the value function $J_t$ is maximized over the time interval $[t,T]$. Note that $J_t = J(\theta_t, S_t, X_t, t)$ depends on $\theta_t$. The pure-strategy Nash-equilibrium point $\theta^*$ should satisfy the following two conditions:
\begin{itemize}
	\item Fix a time $t$. For any time $s\in(t,T]$, player standing at time $s$ will react with $\theta^*$.
	\item Given the objective function $J_t$ and the wealth constraint, $\theta^*$ should also be the optimal strategy for the player standing at $t$.
\end{itemize}
Hence, the equilibrium point as described above should satisfy the HJB equation \ref{hjb1}, and coincides with our optimal strategy which takes the explicit form \ref{explicitpolicy}.

\section{Analysis of Optimal Terminal Wealth}
Having obtained an explicit form of optimal strategy, we would like to explore more intuition behind the DMVO. The way we achieve this is to derive expressions for the optimal terminal wealth and value function. Before going there, we firstly introduce the following notation and result:
\begin{itemize}
	\item Let $\tilde{w}_t$ to be an independent Brownian motion of $w_t$, $\rho$ be the correlation between $w_t$ and $w_{Xt}$. Then we can write:
	$$d\tilde{w}_t = \frac{1}{\sqrt{1-\rho^2}}(dw_{Xt}-\rho dw_t)$$
	\item Substituting the expression of $\theta^*$ in \ref{explicitpolicy} into \ref{ito1} and applying Ito's Lemma, we have:
	\begin{align}
	d(W_t e^{r(T-t)}) = -df_t + \frac{\mu_t - r}{\gamma \sigma_t} dw_t + \sqrt{1-\rho^2}\nu_t \frac{\partial f_t}{\partial X_t} d\tilde{w}_t \label{ito5}
	\end{align}
\end{itemize}
Integrating \ref{ito5} from t to T, we can obtain the optimal wealth. Its mean, variance, and the value function $J_t$ can be derived accordingly. We summarize the result as the following proposition:\\
\begin{prop}
	With a DMVO objective, the terminal wealth, its mean, variance, and the value function are given by:
	\begin{align}
	W_T^* &=W_te^{r(T-t)} + f_t + \frac{1}{\gamma}\int_{t}^{T}\frac{\mu_s - r}{\sigma_s} dw_s + \sqrt{1-\rho^2}\int_{t}^{T} \nu_s \frac{\partial f_s}{\partial X_s}d\tilde{w}_s\\
	var_t[W_T^*]&=\frac{1}{\gamma^2} E_t [ \int_{t}^{T} \frac{1}{\gamma}(\frac{\mu_s - r}{\sigma_s})^2 ds ] + (1-\rho^2) E_t[\nu_s^2 (\frac{\partial f_s}{\partial X_s})^2ds]\\
	E_t[W_T^*] &= W_t e^{r(T-t)} + f_t \\
	J_t &=W_t e^{r(T-t)} + f_t -\frac{1}{2\gamma}E_t[\int_{t}^{T} \frac{1}{\gamma}(\frac{\mu_s - r}{\sigma_s})^2 ds] -\frac{\gamma}{2}(1-\rho^2) E_t[\nu_s^2 (\frac{\partial f_s}{\partial X_s})^2ds]
	\end{align}
\end{prop}
The optimal wealth is represented by two riskless terms $W_te^{r(T-t)} + f_t $ and two risky terms $ \frac{1}{\gamma}\int_{t}^{T}\frac{\mu_s - r}{\sigma_s} dw_s+\sqrt{1-\rho^2}\int_{t}^{T} \nu_s \frac{\partial f_s}{\partial X_s}d\tilde{w}_s$. Riskless terms are driven by anticipated portfolio gains. Risky terms are generated from a hedgeable source $\frac{1}{\gamma}\int_{t}^{T}\frac{\mu_s - r}{\sigma_s} dw_s$ and an unhedgeable source $\sqrt{1-\rho^2}\int_{t}^{T} \nu_s \frac{\partial f_s}{\partial X_s}d\tilde{w}_s$, which is due to the market incompleteness and co-movement between stock and state variable. In addition, they are also driving factors for the variance. The hedgeble term $\frac{1}{\gamma^2} E_t [ \int_{t}^{T} \frac{1}{\gamma}(\frac{\mu_s - r}{\sigma_s})^2 ds ]$ in variance is totally dependent on the state variable. For the second term, the market incompleteness $\rho^2$ is the main influencer. We are confident that an incomplete market gives a more volatile optimal terminal wealth compared to that in a complete market. However, if completeness $\rho^2$ is close between two markets, this deduction may not be true because the sensitivity of $f_t$ to the state variable $S_t$ is also affected by $\rho^2$, where the relationship is convoluted, and conclusion varies on the different economic setting.\\\\
Furthermore, the market incompleteness $\rho^2$ also affects the mean optimal wealth. As discussed before, a low $|\rho|$ leads to a low hedging demand. Specifically, the investor's positive hedging demand will be lower for higher levers of market incompleteness, which results in a lower expected terminal wealth.\\\\
Lastly, findings for the value function are not deterministic. When hedging demand is positive, expected optimal wealth is higher, and the variance is lower in a complete market. However, the relationship is vague when hedging demand is negative, since lower expected optimal wealth will ruin  the lower variance. 
\section{Precommitment Strategy}
At the beginning of the chapter, we addressed that a problem with DMVO objective is time-inconsistent since the investor has incentives to deviate from the optimal policy in the future. Later, we derived a time-consistent investment policy, which both maximizes the DMVO objective within an instant period and investor's motive to change the policy in later time. We now focus on the second method to address the time-inconsistency, namely, precommitment.\\\\
Investor who follows a precommitment strategy maximize her value function at time 0, and she is not allowed to make adjustment to it in the future. Furthermore, if we restrict the investor to stick to her initial policy during the investment horizon (for example, some covenant set by creditors), those who choose the time-consistent strategy will not adjust their policies, and hence, precommitment investors own the advantage in terms of the initial value function. It is also worthy to mention that analytical solutions of problems with DMVO objective have only been computed under precommitment situations. Based on these reasons, we would consider precommitment strategy as a good benchmark compared to the time-consistent strategy.\\\\
It is generally known that precommitment strategy in incomplete markets is a convoluted problem. Hence, we will demonstrate the strategy under a complete market setting. We first define a price density process $\xi_t$, which represents the Arrow-Debreu price per unit probability $P$ of one unit of wealth in some state at time $t$:
\begin{align}
\xi_t = \xi_0 e^{-rt - \frac{1}{2} \int_{0}^{t} (\frac{\mu_s - r}{\sigma_s})^2ds - \int_{0}^{t}\frac{\mu_s - r}{\sigma_s}dw_s}
\end{align}
This is not an unfamiliar result, as we recall that in the hedge neutral measure $P^*$, the Radon-Nikodym
derivative is given by $e^{- \frac{1}{2} \int_{0}^{t} (\frac{\mu_s - r}{\sigma_s})^2ds - \int_{0}^{t}\frac{\mu_s - r}{\sigma_s}dw_s}$. The DMVO objective under a precommitment strategy is formulated as:
\begin{align}
\max_{W_T} E_0[W_T] - \frac{\gamma}{2} var_0[W_T]\\
s.t. \quad E_0[\xi_T W_T] \leq W_0 \label{constraintprecommitment}
\end{align}
In order to obtain an analytical solution, we derive the Lagrangnian:
\begin{align}
L = E_0[W_T] - \frac{\gamma}{2} var_0[W_T] - \lambda (E_0[\xi_T W_T]  - W_0)
\end{align}
The first-order condition is then given by:
\begin{align}
1-\gamma \hat{W_T} + \gamma E_0[\hat{W_T}] - \lambda \xi_T = 0
\end{align}
Noticing properties $E_0[\xi_T e^{rT}] = 1$, $E_0'[W_T] = 1$, and taking $\xi_0=1$, we obtain:
\begin{align}
\hat{W_T} = \frac{1}{\gamma} (1+\gamma E_t[\hat{W_T}] - \xi_T e^{rT}) \label{preprecommitment}
\end{align}
Substituting \ref{preprecommitment} to equality the constraint \ref{constraintprecommitment}, we obtain the analytical solution, which is presented by the following proposition:\\
\begin{prop}
	The optimal terminal wealth of a mean-variance optimizer under precommitment is given by:
	\begin{align}
	\hat{W_T} = W_0e^{rT} + \frac{1}{\gamma} E_0[\xi_T^2]e^{2rT} - \frac{1}{\gamma} \xi_T e^{rT} \label{wealthprecommitment}
	\end{align}
\end{prop}
It is worthy noting the optimal terminal wealth given by the time-consistent strategy is given below, rewritten in terms of the state price density:
\begin{align}
W_T^* = &W_0 e^{rT} + \frac{1}{\gamma} E_0[\xi_T e^{rT} \int_{0}^{T} (\frac{\mu_s - r}{\sigma_s})^2] - \frac{1}{\gamma}[ln\xi_T + rT + \frac{1}{2} \int_{0}^{T}(\frac{\mu_s - r}{\sigma_s})^2 ds]  \label{wealthtimeconsistent}
\end{align}
Results given by two strategies are generally different, as shown above. However, one can proof that for a short decision interval, the wealth under pre-commitment setting given in \ref{wealthprecommitment} is a second-order approximation of that given by time-consistent optimal terminal wealth \ref{wealthtimeconsistent}. Moreover, the market price of risk or the Sharpe ratio, $\dfrac{\mu_s - r}{\sigma_s}$ serves as a comparison measure here. We notice that two wealth coincide when the Sharpe ratio equates to zero. Furthermore, if the Sharpe ratio is set to a constant, optimal wealths from two strategies become more expressive and informative. We summarize the result in the following proposition:\\
\begin{prop}
	Assume the stock has a constant market price of risk, that is, $\dfrac{\mu_s - r}{\sigma_s}=\dfrac{\mu - r}{\sigma}$, the investor's optimal wealth under precommitment strategy and time-consistent strategy, $\hat{W_T}$ and $W^*_T$, are given by:
	\begin{align}
	 \hat{W_T} = &W_0e^{rT} + \frac{1}{\gamma} e^{(\frac{\mu - r}{\sigma})^2T} - \frac{1}{\gamma} \xi_T e^{rT}\nonumber \\
	= &W_0e^{rT} + \frac{1}{\gamma} e^{(\frac{\mu - r}{\sigma})^2T} - \frac{1}{\gamma} e^{- \frac{1}{2}(\frac{\mu - r}{\sigma})^2T - \frac{\mu - r}{\sigma}w_T}\\ \nonumber\\
	W_T^* = &W_0 e^{rT} + \frac{1}{\gamma}  (\frac{\mu - r}{\sigma})^2T
	- \frac{1}{\gamma}[ln\xi_T + rT + \frac{1}{2} (\frac{\mu - r}{\sigma})^2 T]\nonumber\\
	= & W_0 e^{rT} + \frac{1}{\gamma}  (\frac{\mu - r}{\sigma})^2T
	-  \frac{1}{\gamma}(\frac{\mu - r}{\sigma})^2 w_T
	\end{align}
\end{prop}
By some simple algebra, one can easily prove that $\hat{W_T} > W_T^*$, which is a desired result as described at the beginning of the section. Note that in order to compute the numerical result, one can apply either parametric approach (for example, Breeden and Litzenberger (1978) \cite{breedenlitzenberger}) or non-parametric approach (Ait-Sahalia and Lo, 1998 \cite{aitsahalialo}) to estimate price density $\xi$ from a market source.

\newpage
\chapter{Applications and Practical Results}
In this chapter, we illustrate the usefulness of the strategy and theoritical results obtained in previous chapters by applying them in real market situations. Since the discussion in Chapter 3 is based on a market which contains only one risky stock, we will first derive explicit forms for a multiple stock market. Then, in Section 4.2, we study the time-consistent strategy in an economic environment where stocks are driven by the constant elasticity of variance (CEV) model. Lastly, we present the practical performance of different strategies in Section 4.3

\section{Multiple-stock formulation}
We would like to extend the findings in Chaper 3 to a multivariable case in this section. Instead of two correlated Brownian variables, the market with $N$ risky assets and $K$ state variables has uncertainties generated by two vectors of Brownian motions $w = (w_1, ..., w_N)^T$, $w_x = (w_{X1}, ..., w_{XK})^T$ with $N \times K$ correlation matrix $\rho = (\rho_{nm})$. Each element $\rho_{nm}$ denotes the correlation coefficient between $w_n$ and $w_{Xm}$ where $ 1\leq n\leq N, 1\leq m \leq K$. We also denote $\mu = (\mu_1, ..., \mu_N)^T$ as the vector of expect returns, and $\sigma = (\sigma_1, ..., \sigma_N)^T$ as the covariance matrix, where $\sigma_i = (\sigma_{i1},...,\sigma_{iN})$ represents the covariance between ith stock and jth Brownian motion $w_j$. The investor can also choose to invest in one riskless bond with a constant interest rate $r$. Then, the stocks $S = (S_1, ..., S_N)^T$ follow processes below:
\begin{align}
\frac{dS_{it}}{S_{it}} = \mu_i(S_t,X_t,t )dt + \sigma_t(S_t,X_t,t)^T dw_t
\end{align}
for all $ 1\leq i\leq N$. Similarly, the state variables follows:
\begin{align}
dX_{jt} = m_j(X_t, t)dt + \nu_t (X_t,t)^T dw_{Xt}
\end{align}
for all $ 1\leq j\leq K$, and $m = (m_1,...,m_K)^T$, $\nu=(\nu_1,...,\nu_K)^T$ are state characteristics defined similarly as $\mu$. With above notations, the wealth process is then given by: 
\begin{align}
dW_t = [rW_t + \theta_t^T(\mu_t-r)]dt+\theta_t^T\sigma_t dw_t
\end{align}
where $\theta_t = (\theta_{1t},...,\theta_{Nt})$ is the vector of money invested in each stock at time $t$. An investor who follows a DMVO objective aim to solve the problem below:
\begin{align}
&\max_{\theta} E[W_T]-\frac{\gamma}{2}var[W_T] \\
s.t.\quad dW_t &= [rW_t + \theta_t^T(\mu_t-r)]dt+\theta_t^T\sigma_t dw_t \nonumber
\end{align}
The objective shown above corresponds to that in a single stock case. One can formulate a adjusted value function and the HJB equation and derive the optimal strategy in terms of the anticipated gain. We summarize the result in the proposition below, which is, in fact, a generalized version of Proposition 2: \\
\begin{prop}
	The optimal time-consistent investment policy in a multiple stock market is given by:
	\begin{align}
	\theta_t^* = &\frac{1}{\gamma } (\sigma_t \sigma_t^T)^{-1} (\mu_t-r)e^{-r(T-t)} \nonumber\\
	&- (diag(S_t) (\frac{\partial f_t}{\partial S_t})^T+(\nu_t \rho^T \sigma_t^{-1}) (\frac{\partial f_t}{\partial S_t})^T)e^{-r(T-t)}
	\end{align}
	where the anticipated portfolio gain $f_t$ is given by:
	\begin{align}
	f(S_t, X_t, t) = E_t^*[\int_{t}^{T} \frac{1}{\gamma}(\mu_s - r)^T (\sigma_s \sigma_s^T)^{-1}(\mu_s - r) ds]
	\end{align}
	where $E_t^*[\cdot]$ is the expectation under the hedge neutral $P^*$. Two standard Brownian motions with the correlation matrix $\rho$ are given by:
	\begin{align}
	dw_t^* =& dw_t + \sigma^{-1} (\mu_t - r)dt\\
	dw_{Xt}^* =& dw_{Xt} + \rho^T \sigma_t^{-1}(\mu_t - r)dt
	\end{align}
	and the Radon-Nikodym derivative from $P$ to $P^*$ is given by:
	\begin{align}
	\frac{dP^*}{dP} = e^{-\frac{1}{2}\int_{0}^{T}(\mu_s -r)^T(\sigma_s \sigma_s^T)^{-1}(\mu_s -r)ds - \int_{0}^{T}(\sigma_s^{-1}(\mu_s - r))^Tdw_s}
	\end{align}
\end{prop}
It is not suprising that the optimal policy consists one myopic term and one hedging term. Besides, we should also notice that the effect of cross-correlations enters through the second term of hedging demand. Applying similar techniques used in Chapter 3, we can also obtain the optimal terminal wealth:
\begin{align}
W_T^* =&W_te^{r(T-t)} + f_t \nonumber \\
&+ \frac{1}{\gamma}\int_{t}^{T} (\sigma_s^{-1})^T(\mu_s - r) dw_s + \sqrt{I-\rho \rho^T}\int_{t}^{T} \nu_s (\frac{\partial f_s}{\partial X_s})^Td\tilde{w}_s
\end{align}

\section{Constant Elasticity of Variance (CEV)}
We first address a single-stock market where the uncertainty is governed by the CEV process in this section. There are different representations among various literature. Here, under the similar market setup in Chapter 3, the stock shall follow the CEV process below:
\begin{align}
\dfrac{dS_t}{S_t} = \mu dt + \bar{\sigma} S_t ^ {\alpha/2} dw_t
\end{align}
where $\alpha$ denotes the elasticity of instantaneous stock return. It is worthy to note that the variance term now becomes $\bar{\sigma}^2 S_t^\alpha$, and the CEV process is reduced to geometric Brownian motion when $\alpha = 0$. The process has been well studied in the option pricing theory, especially when researchers want to model stock prices with heavy tails. An interesting property is that stock prices approximately follow a non-central Chi-Square distribution (Lindsay \& Brecher, 2010) \cite{lindsaybrecher}. Moreover, the stock price has a heavy left tail when $\alpha < 0$, and has a heavy right tail when $\alpha > 0$ (Cox, 1996) \cite{cox}. It is an important result that we use to analyse the performance of strategies in section 4.3. With the DMVO objective, one can derive an explicit form of the optimal strategy, given in the proposition below. The proof is omitted here since it is not our primary interest in the section. \\
\begin{prop}
	The optimal time-consistent policy under a CEV model is given by:
	\begin{align}
	\theta_t^* = \frac{\mu_t-r}{\gamma \bar{\sigma^2} S_t^\alpha}e^{-r(T-t)} - \frac{1}{\gamma} ( \frac{\mu_t-r}{\bar{\sigma} S_t^{\alpha/2}})^2  \frac{e^{-\alpha r (T-t)} - 1}{r} e^{-r(T-t)}
	\end{align}
	If the market consists of multiple stocks, the optimal policy will take the form:
	\begin{align}
	\theta_t^* = &\frac{1}{\gamma } (\sigma_t \sigma_t^T)^{-1} [(\mu_t-r)\oslash S_t^\alpha ]e^{-r(T-t)}- \frac{1}{\gamma} (\sigma_t \sigma_t^T)^{-1}  [(\mu_t-r)^2 \oslash S_t^\alpha]e^{-r(T-t)}
	\end{align}
\end{prop}
where $\oslash$ denotes the Hadamard element-wise division. From the result above, we note that second term, which is interpreted as the hedging demand, is greatly affected by the elasticity $\alpha$. In addition, based on the term $e^{-\alpha r (T-t)} - 1$, a positive elasticity leads to a positive hedging demand. There is also a qualitative explanation: A positive elasticity implies that the Sharpe ratio decreases with an increasing stock price, which then leads to a negative correlation between stock return and anticipated portfolio gain. Using the similar reasoning in Chapter 3, the negative correlation then gives a positive hedging demand.

\section{Practical Results}
In this section, we would like to prove the usefulness of strategies discussed above from a practical point of view. We have chosen two different sources: one is the real market data, and the other one is simulated stock data. Since the full result requires some parameter estimation involving complicated Monte-Carlo procedures, we reduce them to constant parameters. Because the wealth does not affect our investment policy, which represents money invested in stocks, it is not meaningful to discuss the absolute wealth of strategies. Instead, we evaluate the strategies based on the trend. 

\subsection{Description of Data}
From the consideration of both long term and low volatility, we chose to select weekly stock price from a 10 year horizon. Two sets of data are prepared for the practical use. The first one is the market consists of top 20 constituents from S\&P 500. Stocks which does not have a life of 10 years are abondoned, for example, Facebook. All data are download from Yahoo Finance. \\\\
In order to have prove the usefulness of strategies, we also generated stocks under different economic conditions. Data is generated by constant mean and covariance factors, one from geometric Brownian motion (GBM) and another one from CEV. The summary of data goes below:
	\begin{itemize}
		\item Real data: 20 top constituents of S\&P 500. 10 year weekly stock prices from 2007/10/29 to 2017/11/1 (523 time stamps).
		\item Simulated data: Two subsets of simulated data, from GBM and CEV respectively. 10 year weekly price, 50 stocks for each subset. Mean return is set to 12.5\%. Correlation and variance rate are 0.05 and 0.2 respectively.
	\end{itemize}

\subsection{Design and Methodology}
In order to translate the theoritical investment strategy to computer codes, a few simplification steps need to be taken.\\\\
Firstly, instead of a continuous investment horizon $[0,T]$, our implementation only takes data from discrete time. Choosing weekly data ensures that the data contains sufficient information, and each instant investment interval ($1/52$) is small enough. In our case, each time $t$ is represented by $i/52$, where $1\leq i \leq 523$ denotes the index of the entry throughout the 10-year (523-week) horizon. \\\\
Secondly, we need to know market characteristics such as $\mu_t$ and $\sigma_t$ in order to compute the value of the strategy. It is possible to perform parameter estimation techniques including maximum likelihood and Monte-Carlo simulation. However, an accurate estimation of market parameters requires a system of model building, which deviates the purpose of our topic. Hence, we assume all parameters appeared in our models are constant. Note that this simplification also avoids problems related to partial derivatives.\\\\
Thirdly, we need to set a scheme for the constant mean and covariance estimation. The method we chose is the overlapping batch method. At each time $t>26$, we select the data from previous 0.5 year (26 weeks) as a batch. Sample mean and covariance matrix from that batch are then calculated, which is used as the parameters to calculate the optimal strategies. We also set the return of the riskless bond to 2.5\%, the rate of the 1-year US treasury bill.\\\\
Lastly, after obtaining the mean and covariance of returns, we simulate the investment process at each time stamp $t$. Since wealth does not affect our investment decisions, without loss of generality, we let the initial wealth be zero. At each time, we calculate the optimal policy, and convert it in both money terms and share terms. The sum of money invested in each stock is balanced by the account of the riskless bond. In other words, the profit of the investment is put into the bond account, and the loss is compensated by the bond (the short selling is allowed if calculated policy gives a negative number). For each interval $[t,t+1/52]$, the profit and loss are calculated using the share terms for each stock. We record the total wealth (bond account + stock account) at each $t$, and present the performance by plotting it. 

\subsection{Result and Discussion}
\begin{figure}[h]
	\centering
	\includegraphics[scale=0.8]{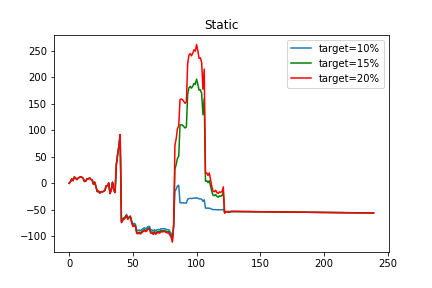}
	\caption{Performance of the static MVO strategy, with different target return}
	\label{staticresult}
\end{figure}

We begin the section with the static MVO strategy. Note that that the result $\omega$ represents the proportion of investment to each stock, hence it is necessary to maintain a variable of investment amount. Apply the simulation steps in the previous subsection, we obtain the result shown in \ref{staticresult}\\

We have chosen three different target returns, namely 10\%, 15\% and 20\%, as shown in \ref{staticresult}. From the curve, we observe that the higher the target return is, the more volatile the wealth is, which is indeed a conclusion that agrees to the theoritical result. It seems abnormal for the peaks in all curves around time 100, but after investigating the investment around the time, we found it is due to the large proportion of investment in Google, which had a significant surge at the same period (around October 2010) when they announced the successful experiment of unmanned vehicles \cite{googleautodrive}. The following plunge is due to the continuous holding of Google (it is a common rule in finance that the price must have a rebound after a continuous surge). Although we did not quit the market at the correct time, this still proved the usefulness of the strategy.\\
\begin{figure}[h]
	\centering
	\includegraphics[scale=0.8]{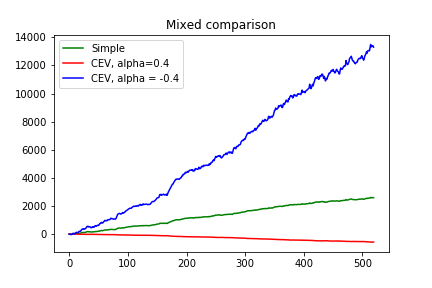}
	\caption{Performance of the DMVO strategy in a GBM economy}
	\label{dynamicresult1}
\end{figure}

\begin{figure}[h]
	\centering
	\includegraphics[scale=0.5]{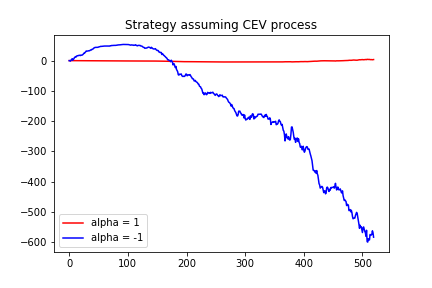}
	\includegraphics[scale=0.5]{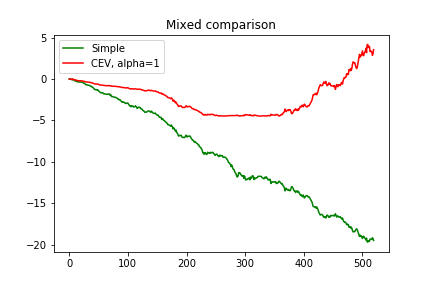}
	\caption{Performance of the DMVO strategy in a CEV economy}
	\label{dynamicresult2}
\end{figure}

\begin{figure}[h]
	\centering
	\includegraphics[scale=0.8]{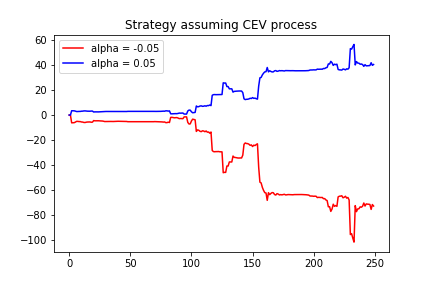}
	\caption{Application of CEV strategy, under real market data}
	\label{CEV1}
\end{figure}

\begin{figure}[h]
	\centering
	\includegraphics[scale=0.8]{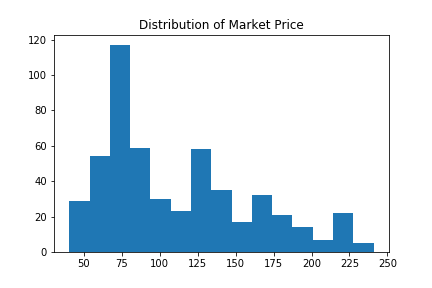}
	\caption{Summary of market index price}
	\label{marketindexhist}
\end{figure}

Next, we would like to explore properties of the dynamic time-consistent strategies. We start with the simulated data. Figure \ref{dynamicresult1} compares performances of three strategies. "Simple" refers to the time-consistent strategy assuming a GBM process. It is clear that the simple strategy beat the CEV strategy with $\alpha = -0.4$. Although the economy is governed by GBM, the CEV strategy with $\alpha = 0.4$ gives the best performance. We interpret this as a result of hedging demand, which is also the reason that drives the $\alpha = -0.4$ case to a negative wealth. In addition, the absolute numbers does not imply the profitability, since they can be scaled by the risk factor $\gamma$, by our design. \\\\
We then focus on the market where stocks are generated by the CEV process with $\alpha = 1$. The result is presented in figure \ref{dynamicresult2}. We split the comparison into two figures because wealth under the CEV with $\alpha = -1$ has such a large scale that make the other two strategies indifferentiable. Overall, figure \ref{dynamicresult2} shows that the strategy assuming CEV with $\alpha = 1$ is the only one that generates positive wealth in the market. Furthermore, U-shape on the red curve proves the effect of hedging demand.\\\\
We now proceed to present some interesting findings regarding the CEV application to the real world. We first show the result of CEV performance with two different $\alpha$ in Figure \ref{CEV1}.
Clearly, performances from two strategies give a similar shape but different direction. We infer from an intuitive rule that the strategy which is closer to market conditions should give a better result, and it is indeed the case for the demonstration here. Recall that stocks which follow the CEV process have the property: The stock price has a heavy left tail when $\alpha < 0$, and has a heavy right tail when $\alpha > 0$ (Cox, 1996) \cite{cox}. We summarize the market index price as the weekly average (assume equal weight for each stock), and plot the histogram in figure \ref{marketindexhist}. Indeed, it shows a distribution with heavy right tail, which explains why CEV with $\alpha = 0.05$ performs better. \\
\begin{figure}
	\centering
	\includegraphics[scale=0.8]{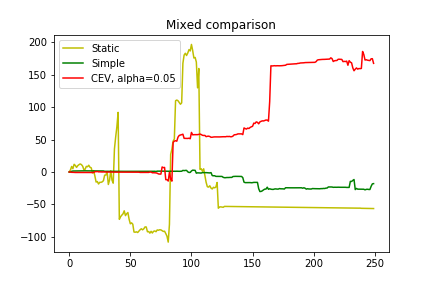}
	\caption{Comparison between dynamic and static, under real market data}
	\label{dynamicvsstatic}
\end{figure}

Lastly, we summarize strategies by plotting static, simple time-consistent and CEV time-consistent on one figure (\ref{dynamicvsstatic}). The CEV strategy with $\alpha = 0.05$ gives the best performance because its assumption is the closet to the market condition, as analyzed above. The simple time-consistent strategy has a stable curve, which is mainly due to maximasation of anticipated portfolio gain (note that the effect of hedging demand is minimal because of model simplification). The static strategy is the most volatile, and it is not suprising because it fully concentrates on the short-term mean return and covariance matrix, which does not consider the hedging issue of the whole investment horizon. \\

Certainly, the result has its limitation: Not only we did simplification to models, but also the data has bias itself. In order to obtain a more robust conclusion, data from different market (outside US) and diversified range (not restricted to top 20) could be chosen for separate experiments.

\newpage

\chapter[Benchmark with Machine Learning]{Benchmark with Machine Learning \footnote{This is an extended topic of the group learning.}}
Recently, machine learning has become a buzz word among the world of technology. Specifically, one branch of it, deep learning, becomes more and more frequently used to detect linear and non-linear relationships. The word "deep" means that many hidden layers of models stacked upon each other, and "learning" connotes the use of neural networks (NN), although can be used with other models. Among various neural network structures, Long short-term memory (LSTM) network, which refers to a recurrent neural network (RNN) composed of LSTM units, is often utilized to handle time series data such as stock data. LSTMs could capture important hidden features from time series and build relational models. The main application of LSTM in finance is to predict the stock price by means of minimising the error between predictions and actual data.\\
\begin{figure}[h]
	\centering
	\includegraphics[scale=0.3]{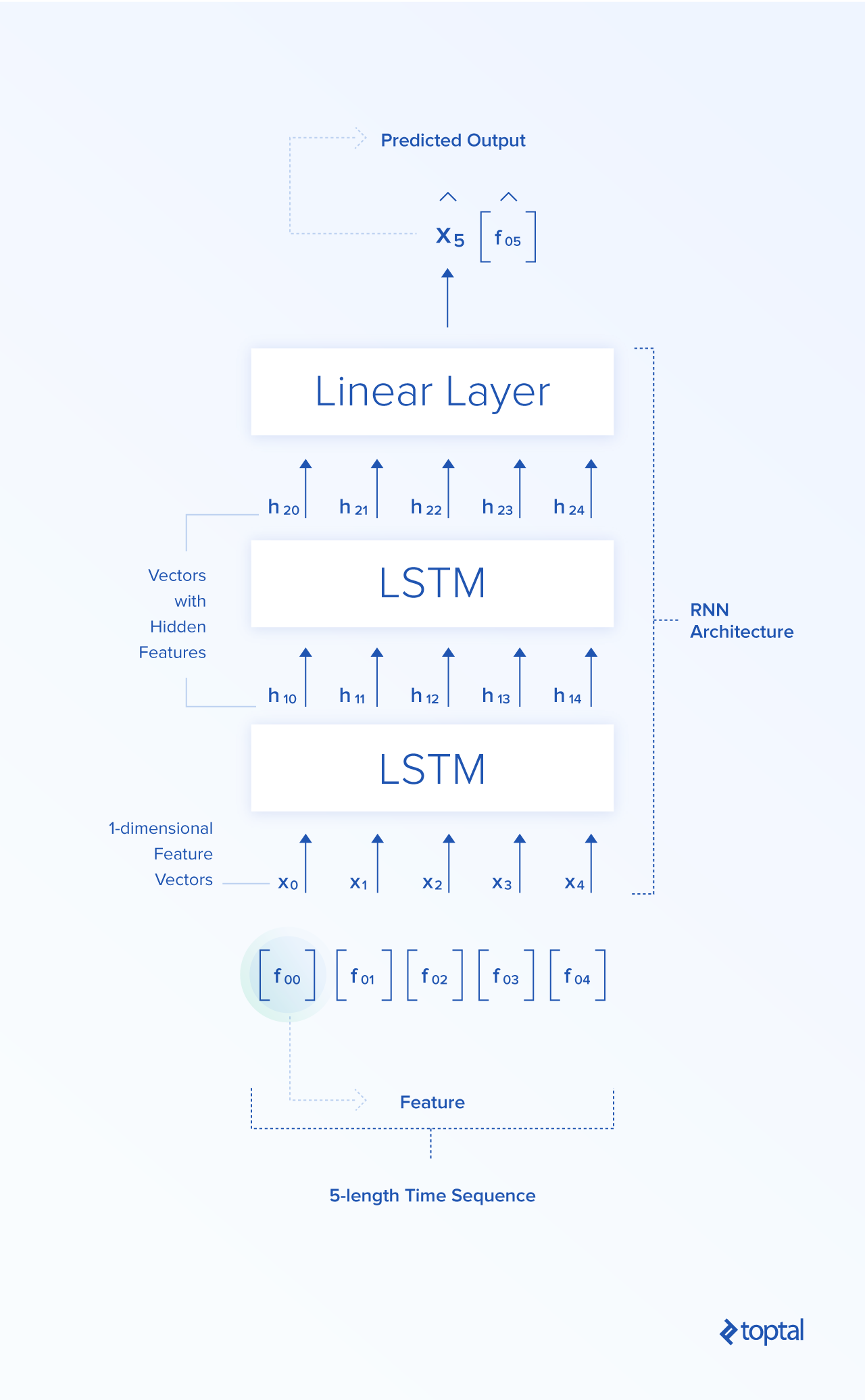}
	\caption{The LSTM structure for time series \cite{lstm}}
	\label{lstm1}
\end{figure}

\begin{figure}[h]
	\centering
	\includegraphics[scale=0.3]{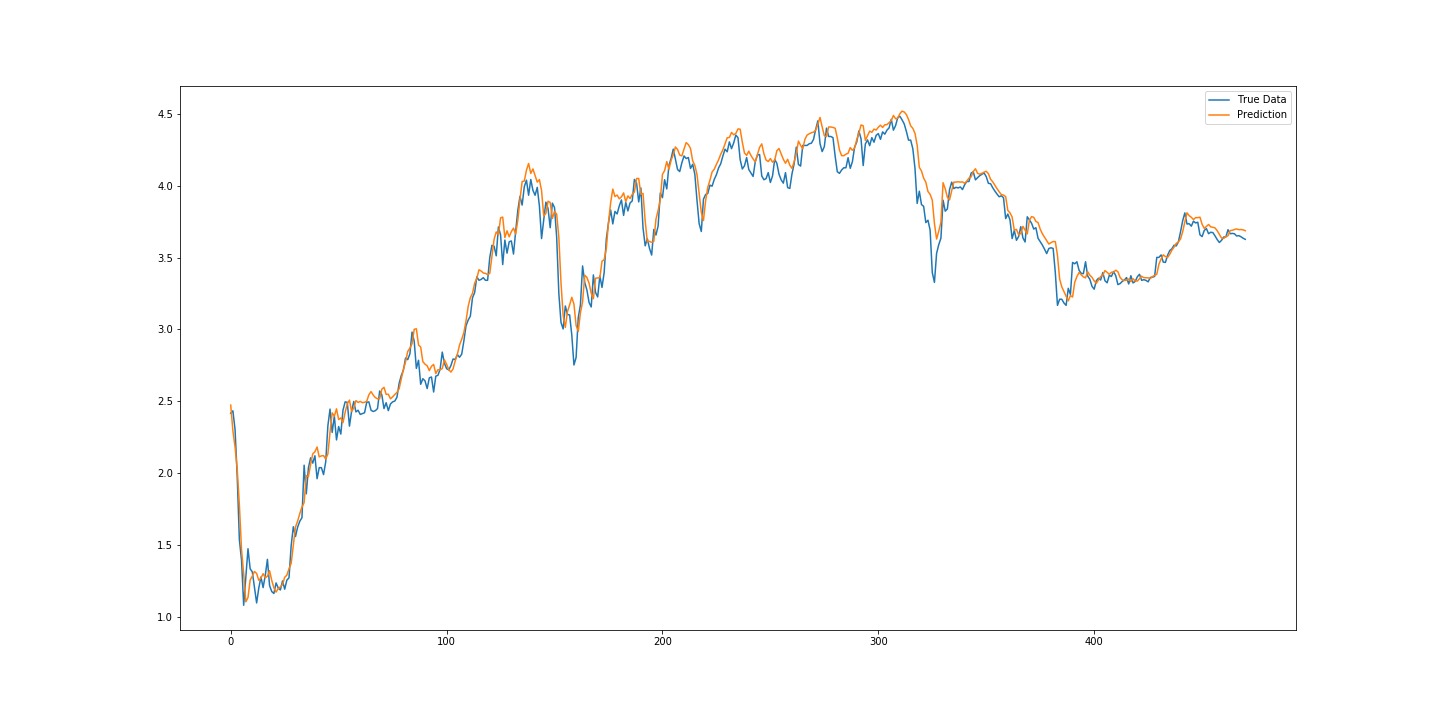}
	\caption{Prediction vs actual data for Keppel Corp}
	\label{lstm2}
\end{figure}

Consider a sample LSTM structure in figure \ref{lstm1}. The LSTM takes a vector of features as its input, for example, five most recent stock prices. In each block, hidden layers with connected neurons and different functional gates are constructed to extract hidden features. All hidden features are finally connected linearly to produce a predicted target. The training procedure is to tune connection parameters such that the error between predicted target and the actual target is minimised, by the method of back-propagation. Figure \ref{lstm2} is a sample output of the LSTM prediction for Keppel Corp (BN4 in Singapore Stock Exchange). It is not surprising to observe the lagging pattern of the prediction curve to the actual curve, since basically, short-term pattern is an important consideration in LSTM. \\

\begin{figure}[h]
	\centering
	\includegraphics[scale=0.8]{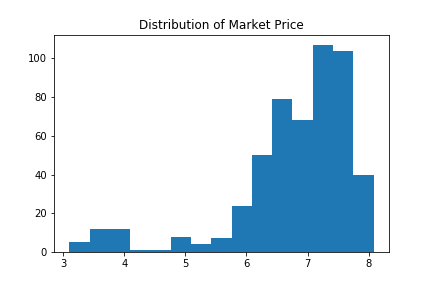}
	\caption{The distribution of selected stocks from SGX}
	\label{pricedist2}
\end{figure}

We would like to compare performance of our dynamic strategy and the LSTM strategy. The LSTM strategy aims to optimize a portfolio under prediction of prices.\footnote{I would like to thank Lim Huan Hock for his kind contribution to LSTM results} The data we used here are 10 year data from 9/9/2007 to 10/9/2017 of 25 stocks listed in Singapore Stock Exchange\footnote{To see the full list, please refer to Appendix A.2}. We observe from figure \ref{pricedist2} that the stock price shows a heavy left tail. Thus, we chose the CEV strategy with a negative $\alpha = -0.1$. Furthemore, since it is hard to compare the wealth curve directly due to different implementation, we decided to compare key statistics of accumulated returns, which serve as good measures for performance. The results are given in table \ref{performancetable}:\\
	
\begin{table}[h]
	\begin{tabular}{|c | c c c|} 
		\hline
		Strategy & Terminal Return & Max Drawdown & Standard Deviation \\ [0.5ex] 
		\hline
		CEV & 47.90\% & -83.82\% & 29.03\% \\  [1ex] 
		\hline
		1 Layer LSTM w. 10 epoches & 20.16\% & -15.23\% & 4.41\% \\  [1ex] 
		\hline
		1 Layer LSTM w. 1000 epoches & -4.42\% & -12.37\% & 2.72\% \\ [1ex] 
		\hline
		2 Layer LSTM w. 10 epoches & 19.89\% & 19.33\% & 5.38\% \\  [1ex] 
		\hline
		2 Layer LSTM w. 1000 epoches & -3.83\% & -11.77\% & 2.68\% \\ [1ex] 
		\hline
	\end{tabular}
\captionof{table}{Performances of strategies}
\label{performancetable}
\end{table}

From the comparsion, we first observe that within the same LSTM structure, the result given by 10 epoch training period has a higher terminal return, max drawdown and standard deviation than those given by 1000 epoches. The main reason is because the more epoches LSTM trained, the closer predictions to lagged prices of the actual data. It is then not difficult to have a more stable return with close price shapes. On the other hand, although there exists the hedging demand, which pulls the return back to a positive number from an intermediate negative return (as indicated by the max drawdown), the CEV strategy is still more volatile, wider ranged than LSTM strategy. After all, it is a strategy that concerns more with a expected future performance rather than recent prices. In conclusion, an extremely risk-averse investor is recommended to manage the investment using LSTM analysis, while dynamic mean-variance strategy brings both high risk and potential high return together.

\appendix
\chapter{List of Selected Stocks}
\section{Selected S\&P 500 Stocks}
1.   Apple Inc.	(AAPL)

2.	Microsoft Corporation (MSFT)

3.	Amazon.com Inc.	(AMZN)

4.	Facebook Inc. Class A (FB) (Not included because of short duration)	

5.	Johnson \& Johnson (JNJ)

6.	Berkshire Hathaway Inc. Class B	(BRK.B)

7.	JPMorgan Chase \& Co. (JPM)

8.	Exxon Mobil Corporation	(XOM)

9.	Alphabet Inc. Class A (GOOGL)

10.	Alphabet Inc. Class C (GOOG)

11.	Bank of America Corporation	(BAC)

12.	Wells Fargo \& Company (WFC)

13.	Procter \& Gamble Company (PG)

14.	Chevron Corporation	(CVX)

15.	Intel Corporation (INTC)

16.	Pfizer Inc. (PFE)

17.	AT\&T Inc. (T)

18.	UnitedHealth Group Incorporated	(UNH)

19.	Visa Inc. Class A (V) 

20.	Citigroup Inc (C)

21. Home Depot Inc. (HD) (Not included because of insufficient data)

22. Verizon Communications Inc. (VZ)

\newpage
\section{Selected Singapore Exchange Stocks}
1. Capitaland (C31)

2. City Developments (C09)

3. ComfortDelgro (C52)

4. DBS Group Holdings (D05)

5. Genting Singapore PLC (G13)

6. Golden Agri-Resources (E5H)

7. Hongkong Land Holdings (H78)

8. Jardine Cycle \& Carriage (C07)

9. Keppel Corp (BN4)

10. Overseas-Chinese Banking Corp (O39)

11. SATS (S58)

12. Sembcorp Industries (U96)

13. Sembcorp Marine (S51)

14. SIA Engineering Company (S59)

15. Singapore Airlines (C6L)

16. Singapore Exchange (S68)

17. Singapore Post (S08)

18. Singapore Press Holdings (T39)

19. Singapore Technologies Engineering(S63)

20. Singapore Telecom (Z74)

21. Starhub (CC3)

22. United Overseas Bank (U11)

23. UOL Group (U14)

24. Wilmar International (F34)

25. Yangzijiang  Shipbuilding  Holdings (BS6)
%
%
\SgIncludeBib{FYP}

\end{document}

%% file: acknowledgments.tex
\thispagestyle{plain}
\begin{center}
	\vspace*{1.5cm}
	{
		\Large \bfseries Acknowledgement
	}
\end{center}

\vspace{0.5cm}

\begin{quote}
	I would like to express my gratitute to people who helped me to make this thesis possible. Firstly, I would like to thank my supervisor Prof. Zhou Chao. Your guidence and patience are the most influential factors that motivated me to conduct the research. I also need to say thanks to my parents, who have been constatntly encouraged me when I feel lost. The complete work will not be there without your continuous support.
\end{quote}
\newpage

%% file: abstract.tex
\begin{abstract}
The portfolio optimisation problem, first raised by Harry Markowitz in 1952, has been a fundamental and central topic to understanding the stock market and making decisions. There has been plenty of works contributing to development of the mean-variance optimisation (MVO) so far. In this paper, one kind of them, namely, dynamic mean-variance optimisation (DMVO) is mainly discussed. One can apply either precommitment or game-theoritical approach to address time-inconsistency in DMVO. We use the second approach to seek for a time-consistent strategy. After obtaining the optimal strategy, we extend the result to a CEV-driven economy. In order to prove the usefulness of them, strategies are fit into both real market data and simulated data. It turns out that the strategy whose assumptions are close to market conditions generally gives a better result. Lastly, a selected strategy is chosen to compare with another strategy came up by deep learning technique.
\end{abstract}